\documentclass[reprint, prb, notitlepage]{revtex4-1}

\usepackage{color}
\usepackage{graphicx}
\usepackage{amssymb, amsmath}
\begin{document}
\title{
Cumulant methods for electron-phonon problems. II. The self-consistent cumulant expansion}
\author{Paul J. Robinson}
\author{Ian S. Dunn}
\affiliation{Department of Chemistry, Columbia University, New York, NY 10027, USA}
\author{David R. Reichman}
\email[]{drr2103@columbia.edu }
\affiliation{Department of Chemistry, Columbia University, New York, NY 10027, USA}
\date{\today}

\begin{abstract}
    In this work we present a self-consistent cumulant expansion (SC-CE) and investigate its
    accuracy for the one-dimensional Holstein model with and without phonon dispersion.
    We show that for finite lattice sizes, the numerical integration of the SC-CE equations becomes unstable at long times.  This defect is partially ameliorated when studying systems in the thermodynamic limit, enabling the demonstration that the SC-CE corrects many deficits of the standard perturbative CE in the (non-dispersive) Holstein model.  The natural phonon damping that arises in the more realistic dispersed Holstein model renders the SC-CE stable, allowing for a complete assessment of the method. Here we find that self-consistency dramatically corrects many of the failures found in the perturbative CE, but also introduces some unphysical features. Finally, we comment on the potential use of SC-CE as a tool for calculating Green's functions in generic many-body problems. 
\end{abstract}

\maketitle

The standard tools of quantum field theory enable the rigorous calculation of Green's functions as an order-by-order expansion in terms of the interaction between excitations.  The use of Dyson's equation and its associated self-energy, along with the Feynman diagram technique, provides a systematic means of partially resumming this infinite order expansion to all orders.
\cite{mahan2013many,fetter2012quantum,abrikosov1963methods}
If the self-energy is expressed in terms of the interacting Green's function, then a self-consistent theory results in which additional diagrams are included.  Perhaps the simplest example of such a self-consistent resummation is the  Hartree-Fock theory for interacting electrons, where the self-energy is calculated to lowest non-trivial order with the fully dressed electron Green's function.  More elaborate resummations also exist which are capable of including effects such as screening beyond the Hartree-Fock level. 

One prominent approximation which includes such screening effects is the GW approach,  which is analogous to Hartree-Fock theory with an RPA-screened interaction.
\cite{Hedin1965, Lundqvist1967SingleparticleSO, Lundqvist1967SingleparticleSOII, Lundqvist1968SingleparticleSO, HybertsenElectron}
The suite of GW approximations also illustrates the distinctions between bare and self-consistent expansions.  If the self-energy is expressed entirely in terms of the bare Green's function, then the so-called $G_{0}W_{0}$ theory results.  This approach has been widely used in the computational condensed matter literature due to its accuracy and affordability.  Unfortunately, for the treatment of realistic systems, the myriad distinct choices for input forms of $G_{0}$ renders the accuracy of this approach sensitive to this input.  In principle this may be solved via a self-consistent solution to the equations.
\cite{Shirley1996, vonBarth1996, Prokofev2008}
However, in addition to the expense of self-consistency, there is no guarantee that the effort is rewarded with increased accuracy. 

For the important benchmark of the electron gas, both the $G_{0}W_{0}$ approximation and fully self-consistent version fail to reproduce important features of the spectral function such as satellite peaks in the spectral function near $k=0$.   To remedy this shortcoming, a cumulant expansion version of the $GW$ approach has been formulated.
\cite{Hedin_1999, almbladh1983beyond}
By construction, the cumulant version of $GW$ is a non-self-consistent approach, formulated by exponentiating the low order $GW$ expansion.  While the cumulant expansion greatly increases the accuracy of the theory with regard to spectral information, it suffers from the same input sensitivity as the standard $G_{0}W_{0}$ approach.  A systematic expansion technique, akin to the standard Dyson expansion with self-consistent resummations, appears lacking in the cumulant Green's function literature.
\footnote{While this work was being written, Ref. \citenum{pandey2022going} appeared, which presents a promising self-consistent approach to electron-phonon problems.
The relationship, if any, to the work presented here remains to be explored.}

In this work we attempt to fill this gap by developing the self-consistent cumulant Green's function theory and test its ability to describe spectral properties in models of an electron coupled to a lattice.
\cite{NeryQuasiparticles}
Such models are of more than circumscribed interest, as they also serve as analogs for purely electronic systems such as the electron gas.  The approach we take was actually formulated nearly five decades ago, but never implemented or studied in any context.
\cite{dunn1975electron}
When compared to the bare cumulant expansion carried out in the previous paper on the same class of models, we find that the simplest self-consistent cumulant theory has the ability to greatly increase the accuracy of spectral features in a non-perturbative manner away from $k=0$, but comes with some deficits, which include ease of solution and the generation of small regions of negative spectral weight.
\cite{RobinsonCumulantI, Gunnarsson1994}
Promisingly, we find that addition of realistic model features such as finite phonon dispersion enhance the stability of the self-consistent equations. Thus we believe the self-consistent cumulant method may be a viable approach to accurately study realistic many-body models in condensed matter physics in a manner independent of input information.

This paper is organized as follows: 
In Sec.~\ref{sec:models}
we introduce the models which are the focus of this study.
In Sec.~\ref{sec:general_derivation} we present a generalized
derivation of the self-consistent cumulant expansion (SC-CE)
for the electron-phonon problem. Some details of the approach are confined to the appendices.
In Sec~\ref{sec:MDIA} we comment on the connection 
between the SC-CE and the Modified Direct Interaction Approximation (MDIA), a classical closure for fluid turbulence. 
In Sec.~\ref{sec:finite_lattice} we study the SC-CE in finite lattice situations, while in Secs. \ref{sec:thermo_non-dispersive} and \ref{sec:thermo_dispersive} we investigate the Holstein
model without and with phonon dispersion in the thermodynamic limit. In the last section we conclude with future perspectives.

\section{Models \label{sec:models}}
We focus here on the Holstein model,\cite{holstein1959studiesI, holstein1959studiesII} and a modified version of the Holstein model
which includes dispersed phonons.\cite{PhysRevB.103.054304}
For both of these models exact VD data exist for comparison.\cite{Bonca2019, PhysRevB.103.054304} 
Work applying the non-self-consistent
CE to other systems is common in the literature,  
both in model systems\cite{mahan1966phonon, dunn1975electron, NeryQuasiparticles, KasCumulant} and, more recently, in ab initio descriptions of realistic materials.\cite{ZhouPredicting}
Self-consistent approaches based on the standard self-energy approach in 
conjunction with the Dyson equation in model\cite{MARSIGLIO1993280,MitricSpectral}
and more elaborate \textit{ab initio} systems have been explored.\cite{RevModPhys.89.015003,PhysRevLett.121.086402,PhysRevB.101.165102}
Relative to this large body of work, self-consistent cumulant approaches have received little to no attention. 
One notable example is the application of a self-consistent
cumulant approach to the electron gas.\cite{PhysRevB.56.12825}
This approach differs from the one we investigate here, and the connections between
and differences of these approaches will be explored in a future publication.
Our study of two well-known models
is presented as a proxy for assessment of the application of such approaches to realistic materials including those with complicated band structure, \textit{ab initio}-determined electron-phonon coupling constants, and multiple phonon modes.

We exclusively consider the case of a single electron in an otherwise empty band coupled linearly
to a bath of phonons
\begin{equation}
    H = H_e + H_p + V.
\end{equation}
The electronic Hamiltonian is given by the tight-binding form
\begin{align}
    H_e &= \sum_k \epsilon_k a_k^\dagger a_k, \\
    \epsilon_k & = -2t_0 \cos k,
\end{align}
and the bath of longitudinal phonons is given by
\begin{align}
    H_p &= \sum_n \omega_0 b_n^\dagger b_n + t_1\left(b_n^{\dagger} b_{n+1} + \text{h.c.}\right)  = \sum_q \omega_q b_k^\dagger b_k, \\
    \omega_q &= \omega_0  + 2  t_1 \cos q.
\end{align}
When $t_1=0$ the model is precisely the standard dispersionless Holstein model.
Here, $a_k (b_k)$ is an annihilation operator for electrons (phonons),
$\epsilon_k$ is the bare electronic energy, and $\omega_q$ is the phonon dispersion.
The electron-phonon interaction is given by
\begin{align}
    V = \frac{g}{\sqrt{N}}\sum_{kq} a_{k+q}^{\dagger}a_{k}\left(b_{q}+b_{-q}^{\dagger}\right).
    \label{eq:epi_hamiltonian}
\end{align}
Throughout this work we characterize our systems with the dimensionless effective electron-phonon coupling strength 
$\lambda=g^2/2t_0\sqrt{\omega_0^2 - 4 t_1^2}$.\cite{MarchandEffect, PhysRevB.103.054304}

\section{Self-Consistent Cumulant Expansion \label{sec:general_derivation}}
The primary quantity of interest in this work is the one-electron (causal) Green's function.\footnote{For single particle models such as ours, this is equivalent to the retarded Green's function.}
In the case of a single electron on the lattice, the thermal trace over all states
is replaced with a trace over zero-electron states
\begin{align}
\mathcal{G}\left(k,t\right) & =-i\Theta\left(t\right)\frac{\text{Tr}\left[e^{-\beta H_{p}}a_{k}(t)a_{k}^{\dagger}(0)\right]}{\text{Tr}\left[e^{-\beta H_{p}}\right]},
\nonumber \\ &
\equiv-i\Theta\left(t\right)\langle a_{k}(t)a_{k}^{\dagger}(0)\rangle.
\label{eq:greens_definition}
\end{align}
We additionally define the spectral function as
\begin{equation}
A(k,\omega) = \pi^{-1}\Im\left[\int_{-\infty}^{\infty}dt \ e^{i\omega t} \mathcal{G}_k(t) e^{-\gamma t}\right],
\end{equation}
including an exponential broadening factor $\gamma$.

In the preceding paper, we followed the standard route of calculating cumulants order-by-order from moments,\cite{mahan2013many}
and we refer the reader to that work for a detailed derivation and discussion of the CE.\cite{RobinsonCumulantI}
This approach, however, does not lend itself to deriving self-consistent methods.
Here, we construct the SC-CE through an alternative derivation first formulated by Dunn,\cite{dunn1975electron} and fully presented in Appendix \ref{app:full_derivation_S}.

\begin{figure*}[!ht]
    \centering
    \includegraphics[width=0.95\textwidth]{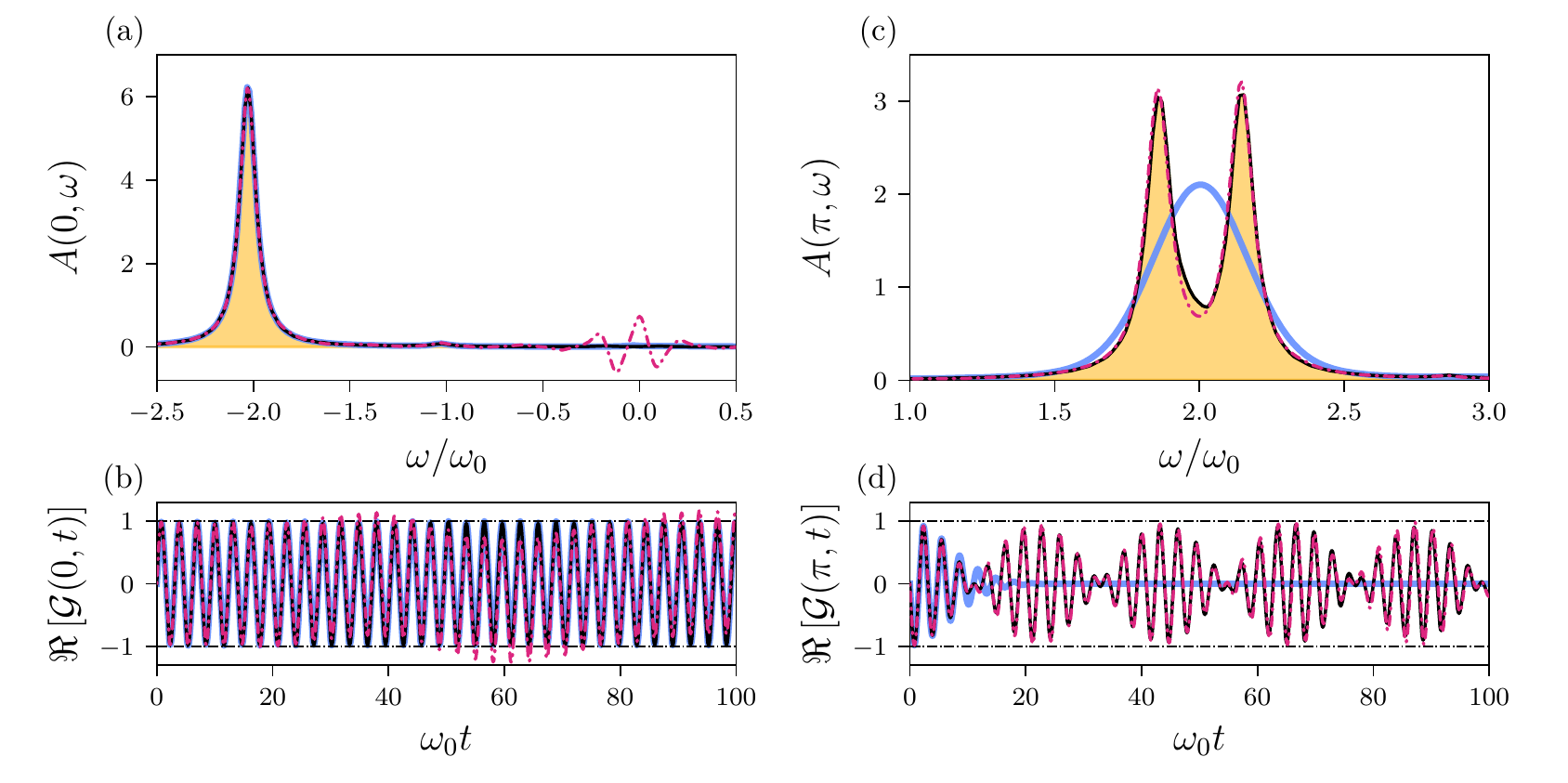}
    \caption{Comparison of the SC-CE (dot-dashed red line), CE (solid blue line) and ED (solid black line with gold shading) for the Holstein model with $N=6$, $\lambda=1/32$ and $T=0.1\omega_0$
    (a) Spectral function for $k=0$. Note the negative spectral weight predicted by the SC-CE.
    (b) $\mathcal{G}(0,t)$ for the three methods at $k=0$.
    (c) The spectral function at $k=\pi$, note that the SC-CE improves upon CE by correctly predicting a large peak splitting.
    (d) Demonstration of the fact that the peak splitting predicted by the SC-CE is a product of recurrences in  $\mathcal{G}(\pi,t)$ which the CE fails to capture.
    The spectra in (a) and (c) are calculated with a broadening of $\gamma=0.05$.
    }
    \label{fig:six_site_weak_coupling}
\end{figure*}

Starting with the time-ordered exponential form of Eq.~\ref{eq:greens_definition},
\begin{align}
    \mathcal{G}_k(t) &= \mathcal{G}_k^{(0)}(t) \langle a_k T e^{-i \int_0^{t} d\tau V(\tau)} a_k^\dagger \rangle, \label{eq:g_t_time_ordered_exp}\\
    \mathcal{G}_k^{(0)}(t) &\equiv -i \Theta(t) e^{-i \epsilon_k t}, 
\end{align}
we directly evaluate the thermal trace, expressed in terms of a differential operator $S$, such that 
\begin{align}
    \mathcal{G}_k\left(t\right) &= \mathcal{G}_k^{(0)}(t) T e^S = -i \Theta(t) e^{C_k(t)}.
\end{align}
Here,
$T$ the time-ordering operator places operators in order of increasing time from right to left,
and 
$C_k(t)$ is  the sum of the cumulants of all orders.
This contrasts with the standard approach, where the thermal trace is evaluated for each order individually
rather than for all orders at once. 
Finding $S$ is 
not practical since determining a closed form for $S$ is tantamount to solving the full problem.

The advantage of introducing $S$ is that it can be transformed to other bases before the expansion 
and resummation.
The details of this transformation are included in Appendix \ref{app:full_derivation_S},
the result of which is that occurrences of $\epsilon_k t$ in the 
integrand of the cumulant can be replaced with a nearly arbitrary function of time and momentum.
The most intriguing use of this transformation is the creation of self-consistent versions of the cumulant expansion.

Dunn explored $S$ in the context of the Fr{\" o}hlich model with what he called the ``improved cumulant'' approach, where one sets $\epsilon_k t \to E_k t$, where $E_k$ is a renormalized energy.\cite{dunn1975electron}
He also suggested, but did not explore, a fully self-consistent version of the cumulant-expansion,
where $i\epsilon_k t \to C_k(t)$.
This formulation is the focus of the work presented here.

To leading order, the resulting self-consistent equation for the cumulant is (see Appendices \ref{app:full_derivation_S} and \ref{app:sc-ce_recursive})
\begin{equation}
\begin{aligned}
        C_k(t) = -i \epsilon_k t 
    - &
    i\frac{g^2}{N}
    \sum_q 
    \int_0^t  d\tau 
    \int_0^{\tau}d\tau' 
    D^{(0)}_q(\tau-\tau') \\
    \times & e^{- C_k(\tau) + C_{k-q}(\tau) - C_{k-q}(\tau') + C_{k}(\tau')}.
\end{aligned}
\label{eq:sc_ce_cumulant}
\end{equation}
Here,  $D^{(0)}_q(t)$ is the free phonon propagator,
$D^{(0)}_q(t) =  -i\left[(1 + N_q) e^{-i\omega_q |t|} + N_q e^{i\omega_q |t|}\right]$, 
and $N_q$ is the Bose occupation factor, $N_q = (\exp(\beta \omega_q) - 1)^{-1}$.
Eq. \ref{eq:sc_ce_cumulant} is a difficult integral equation to solve,
coupling the solution for a fixed wave vector at time $t$ to all other Green's functions for other momenta and all previous times. 
It is, however, readily transformed into a self-consistent 
expression for the interacting Green's functions, which is 
more convenient for calculations
\begin{equation}
\begin{aligned}
\frac{d \mathcal{G}_{\vec{k}}(t)}{d t} & =  
-i\epsilon_k \mathcal{G}_{\vec{k}}(t) 
\\
&
- i
\frac{g^2}{N}
\sum_{\vec{q}}  
\int_0^t d\tau
D^{(0)}_{\vec{q}}(t - \tau)
\frac{
\mathcal{G}_{\vec{k}}(\tau)
\mathcal{G}_{\vec{k}-\vec{q}}(t)}
{\mathcal{G}_{\vec{k}-\vec{q}}(\tau)}.
\end{aligned}
\label{eq:sc_action_gt}
\end{equation}

The properties of the solutions to Eqs.~\ref{eq:sc_ce_cumulant}-\ref{eq:sc_action_gt}
are not readily apparent; however, it can be shown by expressing Eq. \ref{eq:sc_ce_cumulant} recursively 
(Appendix \ref{app:sc-ce_recursive})
that the weak-coupling limit of this SC-CE returns to that of the standard CE.
The recursive SC-CE solution demonstrates that the SC-CE approximates higher-order cumulants by screened second-order cumulants.

The recursive expression Eq.~\ref{eq:sc_action_gt} suggests that SC-CE might suffer from some of the same problems as the fourth-order CE which introduces, e.g., negative spectral weight.
However, it is not clear whether terms which appear to be problematic in the recursive formulation are
suppressed when Eq.~\ref{eq:sc_action_gt} is solved directly. 
Before taking up the question of the performance of the SC-CE, we first take a detour, exploring Eq.~\ref{eq:sc_action_gt} from a different perspective.

\section{Connection to a Theory of Turbulence\label{sec:MDIA}}
In this section we briefly discuss how the formalism developed above and in the appendices is connected to a seemingly unrelated technique in the classical theory of turbulence which has been employed to study the closure problem in passive scalar advection models.  As is well-known, one of the earliest successful self-consistent closures in field theory is Kraichnan's Direct Interaction Approximation (DIA), which when applied to the Navier-Stokes equation or to stochastic models of turbulent flow, makes detailed predictions for a variety of observable quantities.
\cite{leslie1973developments, mccomb1990physics}
In the context of the Holstein model studied here, the DIA is tantamount to a simple one-loop approximation to the self-energy without vertex renormalization, namely the self-consistent Migdal approximation.\cite{Engelsberg1963}

Models of turbulence can also be studied in a systematic way within the Mori-Zwanzig formalism.
\cite{zwanzig2001nonequilibrium}
Vanden-Eijnden and collaborators have demonstrated that for passive scalar models of turbulence, the DIA arises from a simple self-consistent manipulation of the standard time-retarded memory function.
\cite{PhysRevE.58.R5229,kramer2003closure}
However it is well-known that an exact time-local framework also exists for the description of equilibrium and non-equilibrium dynamical phenomena in statistical mechanics.
\cite{Tokuyama1976}
Time-retarded and time-local methods essentially correspond to the resummation of perturbation theory in distinct classes of cumulants, and both approaches are formally exact if all cumulants are included.\cite{Yoon1975}  Appealing to the time-local framework, Vanden-Eijnden and collaborators developed an analog of the DIA that is local in time, which they called the Modified Direct Interaction Approximation (MDIA), and demonstrated that the MDIA was the most accurate simple closure for the passive scalar models of turbulence they studied.
\cite{kramer2003closure}
It is important to note that in the theory of open quantum systems, the time-local approach has been employed for many years to study the relaxation of a quantal system connected to a bath.
\cite{breuer2002theory}
However these calculations are never formulated and employed in a self-consistent manner.

One can heuristically follow the reasoning of Kramer {\em et al.} to see the connection between the MDIA and Eq.~\ref{eq:sc_action_gt}.  Consider the Dyson equation in the time domain for $t>0$, 
\begin{equation}
i\frac{d \mathcal{G}_{\vec{k}}(t)}{d t} - \epsilon_k \mathcal{G}_{\vec{k}}(t)-\int_{0}^{t}d\tau \Sigma(\vec{k},t-\tau)\mathcal{G}_{\vec{k}}(\tau)=0,
\end{equation}
where for the Holstein model in one dimension $\Sigma(\vec{k},t)= i \frac{g^2}{N}
\sum_{\vec{q}} D^{0}_{\vec{q}}(t)\mathcal{G}^{(0)}_{\vec{k}-\vec{q}}(t)$ to lowest order in perturbation theory. 
Next, one may use the relationship $\mathcal{G}^{(0)}_{\vec{k}-\vec{q}}(t-\tau)=-i[\mathcal{G}^{(0)}_{\vec{k}-\vec{q}}(\tau)]^{-1} \cdot\mathcal{G}^{(0)}_{\vec{k}-\vec{q}}(t)$ in the self-energy term.
Finally, partial resummation may be affected by replacing all bare Green's functions with dressed ones, which converts the Dyson equation precisely into Eq.~\ref{eq:sc_action_gt}.  Note that while this provides the lowest order in a series of self-consistent cumulant equations, systematic higher-order terms may be constructed order by order.  Dunn has discussed higher-order terms from the standpoint of the self-consistent expansion detailed in the appendices. At higher orders, the procedure outlined by Dunn and the renormalization technique discussed here will differ. Such higher-order terms are cumbersome and will not be considered further in this work. 

The connection to the classical MDIA provides an interesting perspective on the expected behavior of the self-consistent cumulant expansion.  For example, it was noted in Paper I that inclusion of the 4th order cumulant may improve the accuracy of the approximate Green's function, but comes at the expense of potentially unphysical behavior, such as the generation of regions of negative spectral weight for some wave vectors.\cite{RobinsonCumulantI}  One may ask if this deficit is cured within the lowest order self-consistent version of the theory.  Within the classical theory of turbulence it is well-known that the DIA approximation is ``weakly realizable" in the sense that it is exact for certain well-defined models which preserve some physical symmetries and properties.  The MDIA is not even weakly realizable, and thus despite its superior accuracy, the price to be paid may be some degree of unphysical behavior, such as negative spectral weight.
\cite{eijnden1997contribution}
As we shall see in the following sections, the same is true of the SC-CE for quantum models.

\begin{table}[!t]
    \centering
    \begin{ruledtabular}
    \begin{tabular}{l l l l}
$\lambda$ & $T/\omega_0$ & $\max(|\mathcal{G}(0,t)|)$ & $\max(|\mathcal{G}(\pi,t)|)$ \\
\hline
$1/32$ & $0.10$ & $1.49$ & $1.00$ \\
$1/8$ & $0.10$ & $1.69$ & $1.00$ \\
$1/2$ & $0.10$ & $1.41\times 10^1$ & $5.61\times 10^1$ \\
$1$ & $0.10$ & $2.11 \times 10^4$ & $1.52\times 10^5$ \\
$1/32$ & $1.00$ & $2.01$ & $1.00$ \\
$1/8$ & $1.00$ & $3.06$ & $1.33$ \\
$1/2$ & $1.00$ & $4.23\times 10^4$ & $2.33\times 10^4$ \\
$1$ & $1.00$ & $3.88\times 10^6$ & $8.21\times 10^6$
    \end{tabular}
    \end{ruledtabular}
    \caption{Demonstration of the growth of SC-CE divergences for the six-site Holstein model.
    For weak couplings ($\lambda = 1/32 $ and $1/8$) and $T=0.1\omega_0$ the divergences are suppressed (thought the Green's function does still display non-physical magnitude $>1$). 
    For $T=1\omega_0$, only $\lambda=1/32$ has a small enough magnitude at $t=100$ to be useful for spectral analysis.
    For both temperatures $\lambda=1/2$ and $1$ rapidly diverge. For all calculations, the integration time-step is $0.001\omega_0^{-1}$.}
    \label{tab:six_site_divergence}
\end{table}

\begin{figure}[!t]
    \centering
    \includegraphics[width=0.45\textwidth]{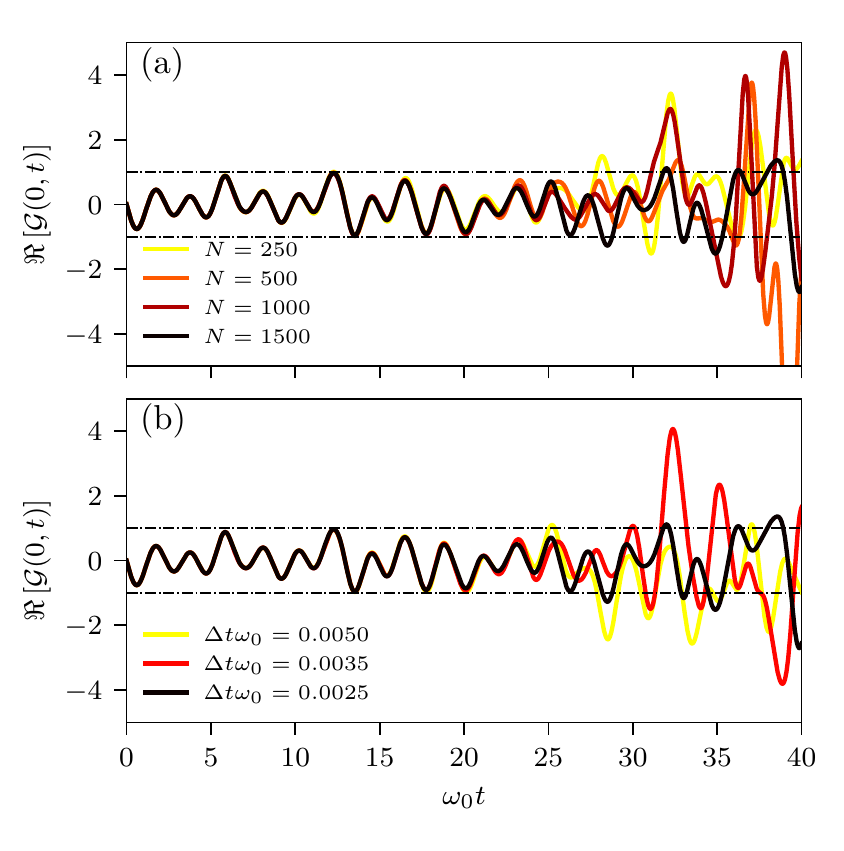}
    \caption{Convergence of the real part of $\mathcal{G}(0,t)$ as calculated by the SC-CE with respect to (a) system size, and (b) time-step. 
    In all cases, $T=0$, $\lambda=\omega_0=1$, and 
    the horizontal dashed-black lines mark $\pm1$ and serve as a guide of when non-physical behaviour sets in.
    In (a) the time-step is uniformly $0.0025\omega_0^{-1}$. $\mathcal{G}(0,t)$ is nearly converged with system size up to approximately $25\omega_0^{-1}$ for $N > 250$, but for longer times the functions diverge. 
    For the case of $N=1500$ (b) demonstrates that very small time steps are necessary to converge the SC-CE at later times, and large time steps worsen the longer-time behaviour. However, continued decrease of the time step cannot control the instabilities at long times within our integration scheme.
    }
    \label{fig:holstein_sc-ce_convergence}
\end{figure}

\section{Results}

For our investigations of both the dispersive and non-dispersive Holstein models, we compare the SC-CE to
exact spectral functions calculated using VD and Exact Diagonalizaton (ED), as well as with the standard CE. 
For the dispersionless Holstein model at intermediate coupling, we compare against the finite-temperature VD data of Bon{\v c}a, Trugman, and Berciu.\cite{Bonca2019}
As in the companion paper, we can compare with these limited sized systems because 
on the scale of the heat maps the spectral functions appear to change only slightly as the system size increases.\cite{RobinsonCumulantI}
For the zero-temperature dispersive Holstein model we compare with the VD data of Bon{\v c}a and Trugman\cite{PhysRevB.103.054304} which are computed in the thermodynamic limit.
We do not review the numerical benchmark approaches here and refer the reader to extensive literature on these methods.\cite{Bonca2019,Trugman1990,Bonca1999,Ku2002,Jaklic2000,Prelovsek2013}
Details of the CE and truncated-ED (``$K$-phonon approximation'') methods are presented in the companion paper.\cite{RobinsonCumulantI}

\subsection{Six-Site Holstein Model\label{sec:finite_lattice}}

We begin with the six-site Holstein model at very weak coupling ($\lambda=1/32$) and low-temperature ($T=0.1\omega_0$).
One may expect
the standard CE to perform well in this parameter regime because the coupling is so weak.
However, the perturbative CE is a short-time expansion
as illustrated in the preceding work.
Thus, there is no guarantee of its success even for very small $g$ away from short times. 
As we will see, even for such small coupling values the standard
second order cumulant fails for $k\neq 0.$

In Fig.~\ref{fig:six_site_weak_coupling}, we compare the spectral functions and $\mathcal{G}(k, t)$ for $k=0$ and $k=\pi$ using SC-CE, CE, and ED at $T=0.1\omega_0$.
The ED results are truncated at the three-phonon level which should be sufficient given the coupling strength.
Fig.~\ref{fig:six_site_weak_coupling}(a) shows that the spectral function at $k=0$ is nearly exactly described by CE (as expected for weak coupling and $k=0$).
The SC-CE predicts the main features approximately as accurately, but inserts a region of non-physical negative spectral weight around $\omega = 0$.

In Fig.~\ref{fig:six_site_weak_coupling}(b), 
the time dependence of
$\mathcal{G}(0, t)$ is plotted, revealing the origin of the negative spectral weight. The SC-CE result deviates from ED via the addition of a long wavelength oscillation peaking for the first time at $\sim 40 \omega_0^{-1}$. 
This additional structure introduces non-physical behaviour into  $\mathcal{G}(0, t)$ with a norm that exceeds unity.

The presence of this negative spectral weight is not surprising
given the understanding of the behaviour of
higher-order cumulants. 
As discussed in the preceding companion paper,
within the context of perturbative higher-order cumulant calculations,
this problem
has been known for some time. \cite{RobinsonCumulantI, Gunnarsson1994,Zuehlsdorff2019, Anda2016}
Additionally, the M-DIA, a classical analogue of the SC-CE for turbulence discussed
in Sec. \ref{sec:MDIA}, can also produce non-realizable negative spectral regions.\cite{eijnden1997contribution}

While for the example of $k=0$ at weak coupling the SC-CE is \textit{worse} than the CE, 
the CE is less accurate away from $k=0$, so we next consider the case of
$k=\pi$ in Fig.~\ref{fig:six_site_weak_coupling}(c).
The most prominent feature of the ED spectrum is a split main peak 
with splitting centered
at $\omega = 2 \omega_0$.
The CE, while generally placing the spectral intensity in the correct position does not include a peak splitting and thus provides a qualitatively incorrect description of the spectrum.
The SC-CE, on the other hand, replicates the ED spectrum nearly exactly, including the prominent peak splitting. 
Fig.~\ref{fig:six_site_weak_coupling}(d) demonstrates that
the SC-CE captures the splitting because it accurately describes the long-time dynamics, specifically recurrences in the temporal evolution of $\mathcal{G}(\pi, t)$.
While the initial damping of $\mathcal{G}(\pi, t)$ is well-described by both CE and SC-CE, the CE does not recur once it decays for the first time at $\sim 10\ \omega_0^{-1}$.
The SC-CE, however, continues to recur periodically, in excellent agreement with the ED result, and without
introducing any non-physical behaviour.
It should be noted that even inclusion of the fourth-order
CE does not capture this behaviour. 
Thus, the SC-CE can describe some non-perturbative effects beyond accessible low-order CEs.

While the SC-CE results for $\lambda=1/32$ and $k=\pi$ demonstrate remarkable improvements over the CE,
only a limited parameter region was accessible for this calculation because the SC-CE rapidly diverges for even slightly larger couplings when $N=6$.
In Table~\ref{tab:six_site_divergence}, we demonstrate that these divergences are exacerbated by increasing coupling strength when $N$ is fixed at a small value, and, to a lesser extent, by increasing temperature. 
We characterize the severity of the divergence with the maximum of $|\mathcal{G}(k, t)|$ over the range $0 \le t\le 40$.  

For $T=0.1\omega_0$ or $1.0\omega_0$ and $\lambda=1/32$ or $1/8$, the Green's function does not dramatically deviate from unity.
Further, for several cases at $k=\pi$ the magnitude does not exceed one within the observed time window.
In these cases, as well as the cases of weak deviations from unity, a modest damping enables us to calculate sharp spectral functions (e.g. Fig. \ref{fig:six_site_weak_coupling}(a) and (c)).
This is not the case for the larger couplings. 
These divergences render Green's functions for finite sized lattices unusable for spectral interpretation.
The origin of these divergences is unclear, but they appear to be part of the SC-CE rather than numerical in nature.
Finally, we note that the dependence of the divergent behaviour on temperature is a weaker effect.

While the divergences may be fundamental to finite sized SC-CE, they are reminiscent of the  divergences we observed in the fourth-order CE, and there we were able to eliminate some of this problematic behaviour by increasing
the system size towards the thermodynamic limit.
Following this line of investigation, we devote the following two subsections
to testing the SC-CE  in systems approaching the thermodynamic limit, 
which is the case of most relevance for the description of experiments.

\subsection{Thermodynamic Limit of the Holstein Model\label{sec:thermo_non-dispersive}}

\begin{figure}
    \centering
    \includegraphics[width=0.5\textwidth]{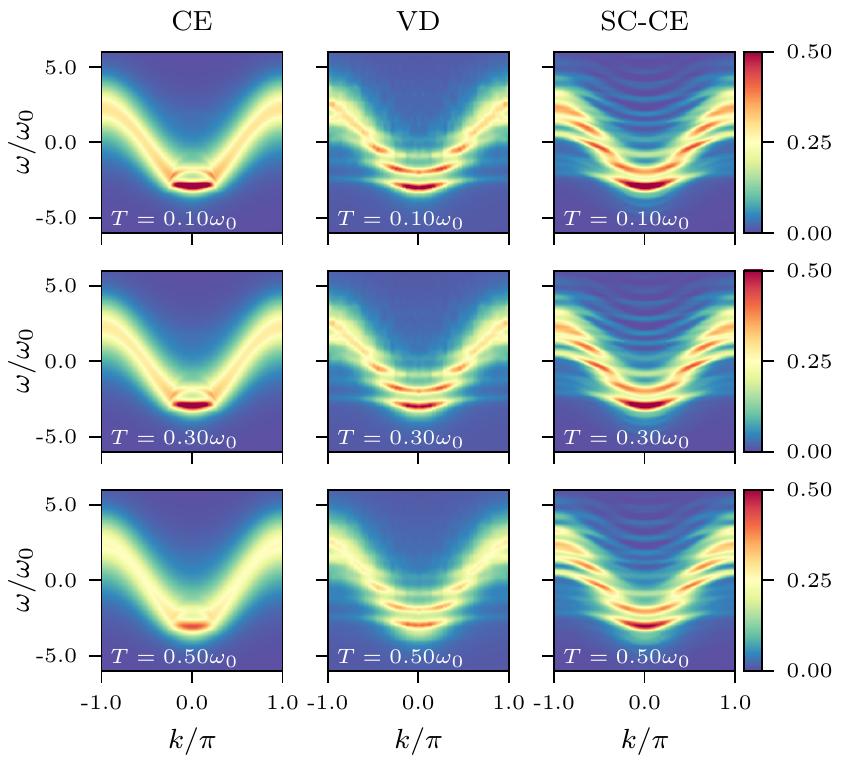}
    \caption{
    Heat maps of the spectral function for the finite temperature Holstein model $(\lambda=\omega_0=1)$.
    Left: the CE with $N=250$. The CE is converged on this scale for this value of $N$.
    Center: VD results of  Bon{\u c}a \textit{et al.}\cite{Bonca2019} for six sites which reasonably approximates the thermodynamic limit on this scale.
    Right: the SC-CE with $N=1500$.
    All three methods use a large broadening of $\gamma=0.25$.
    The SC-CE was calculated with a time-step
    of $0.005 \omega_0^{-1}$ and the Green's function was truncated beyond the first time its norm exceeded unity for each momentum.
    The SC-CE improves upon the CE and semi-quantitatively captures the VD spectra at all temperatures.}
    \label{fig:holstein_heat_maps}
\end{figure}

\begin{figure*}
    \centering
    \includegraphics[width=0.95\textwidth]{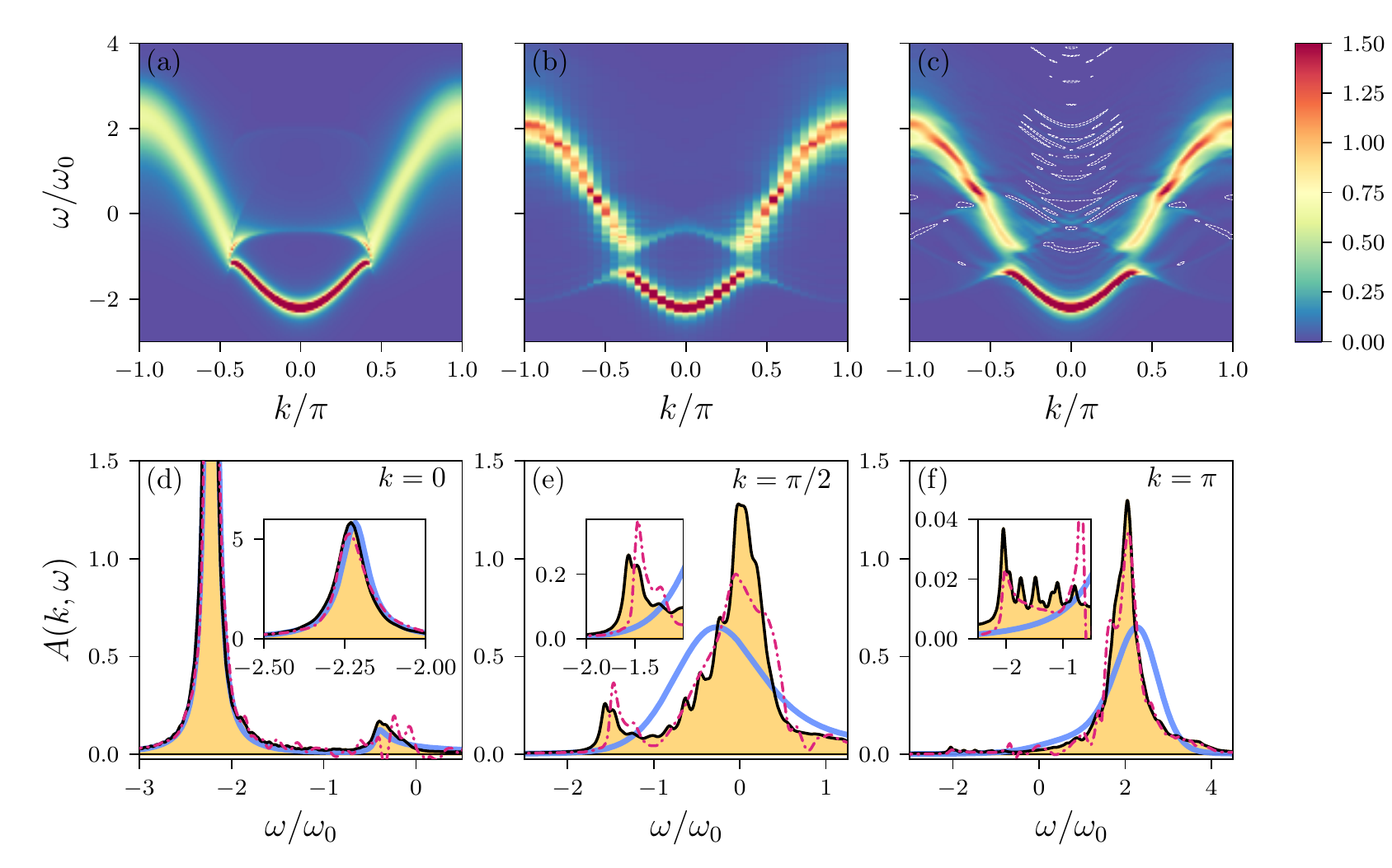}
    \caption{Spectral functions of the dispersive Holstein model for $T=0, \omega_0=t_0=1, \lambda=0.5$ and $t_1=0.4$.
    The comparison between the (a) $N=2000$ CE, (b) VD results of Bon{\v c}a and Trugman,\cite{PhysRevB.103.054304} and (c) $N=2000$ SC-CE demonstrates that the SC-CE accurately captures the VD results across all momenta while CE is accurate only at $k=0$. 
    In (c), small regions of negative spectral weight are bounded with a white dashed line.
    Slices of the spectral functions at $k=0, \pi/2, $ and $\pi$ are presented in (d), (e), and (f) respectively. Each slice compares the CE (solid blue line), VD (solid black line with gold shading),
    and the SC-CE (dot-dashed red line).
    }
    \label{fig:dispersive_heat_map}
\end{figure*}

Motivated by the difficulty of reaching the strong coupling regime in a six-site system, 
we explore the Holstein model with increasingly large numbers of sites as a means of reaching the thermodynamic limit and potentially suppressing instabilities.
In Fig.~\ref{fig:holstein_sc-ce_convergence}(a) we study intermediate coupling $(\lambda = 1)$,
showing that for short times the SC-CE
Green's function does become stable and appears to converge to the infinite lattice limit over some time
scale as $N$ is increased. 
However, beyond $t \approx 30\omega_0^{-1}$ the dynamics 
for different $N$ values diverge dramatically, indicating that convergence has not been reached. 
Further, increasing the system size from $N=1000$ to $N=1500$ does not definitively push back the divergence,
so it remains unclear how effective increasing the system size is at pushing back the divergence. 
Fig.~\ref{fig:holstein_sc-ce_convergence}(b) shows that decreasing the integration time step also eventually becomes ineffective in suppressing instability at long times. 
It remains unclear if the long-time behaviour we observe is an intrinsic failure of the SC-CE
approach, as it is for the fourth-order CE in the same model, or if it is a numerical issue associated with the integration 
of a non-linear integro-differential equation. 
Either way, this behaviour presents a practical limit to how far we can currently converge results for the Holstein model.
If the divergence can be overcome for $N\to\infty$, a different numerical approach which more effectively evaluates the highly-oscillatory momentum integral in Eq.~\ref{eq:sc_action_gt} might provide a better path to convergence.\cite{olver2008numerical, olver2010fast}

By using the times before which $|\mathcal{G}(k,t)| > 1$ and sufficient damping ($\gamma=0.25$), we can
compute spectra from SC-CE in a stable fashion. 
In Fig.~\ref{fig:holstein_heat_maps} 
we compare the spectra of the CE and the SC-CE to the VD results of Bon{\v c}a \textit{et al.}\cite{Bonca2019}
We find that the SC-CE and VD results are generally in agreement over
the whole range of temperatures and momenta.
The SC-CE correctly predicts that the quasiparticle peak flattens and diminishes in intensity
from $k=0$ to $k=\pi$.
This contrasts with the CE which, while accurate for $k=0$, has an abrupt transition where the
quasiparticle peak vanishes and the structured features are replaced by a single broad peak.
We noted in the previous paper that the deficits of the second-order CE could not be generally corrected by inclusion of the fourth-order term.\cite{RobinsonCumulantI}
By contrast, the curvature of the SC-CE satellite peaks is qualitatively correct and the
appearance of the spectrum is correct across the entire range of $k$-space.
Considering that CE predicts qualitatively incorrect satellite curvature, and predicts a nearly featureless spectrum for $k \geq \pi/3$, the SC-CE provides a vast improvement.

Due to the substantial broadening we have used, we cannot make quantitative statements concerning 
the fine structure of the spectral functions.
It does appear, however, that the
SC-CE generally over-weighs the intensity of the vibronic 
peaks.
The SC-CE most resembles VD at lower temperatures, and most severely deviates from exact results
at higher temperatures 
where we have less confidence in our ability to converge the Green's function in this non-dispersive case.

Increasing the system size towards the thermodynamic limit improves the stability 
of the SC-CE, and in light of the excellent spectra produced, the question becomes whether this method could be applicable to realistic systems. 
The simplicity of the Holstein model with its single, non-dispersed, phonon mode also 
potentially contributes to the unstable behaviour, and a more realistic model which
naturally generates more internal damping may lead to better behaviour.
The next subsection addresses this question with an investigation of the dispersive Holstein model.

\subsection{Dispersive Holstein Model\label{sec:thermo_dispersive}}

We now consider a dispersive Holstein model at $T=0$, with  $t_1=0.4, \lambda=0.5$ and $\omega_0 = t_0 = 1$.
As we will see, when naturally dispersive phonons are present, the divergence seen for the SC-CE is strongly suppressed and 
we are able to propagate $\mathcal{G}(k, t)$ to 
long times, allowing for analysis of the fine structure as well as the main spectral peaks.

In Fig. \ref{fig:dispersive_heat_map} we compare (a) the CE, (b) the VD spectra of Bon{\v c}a and Trugman,\cite{PhysRevB.103.054304} 
and (c) the SC-CE.
The SC-CE much more closely resembles the VD result than the CE.
In particular, it correctly predicts the parabolic behaviour of the quasiparticle peak near $k=0$, 
the gap in the spectral function close to $k\approx 0.3 \pi$,
and the loss in intensity in the spectral signatures as $k \rightarrow \pi$. 
Other features, such as the main satellite band, appear better described as well.
For $k=0$, the CE is more accurate than the SC-CE. 
As illustrate in 
Fig. \ref{fig:dispersive_heat_map}(d), the SC-CE introduces noise and 
small regions of negative spectral weight around the first 
satellite and slightly underestimates the quasiparticle weight. 
At intermediate $k$, the SC-CE correctly predicts the formation of a gap between the quasiparticle peak 
and the first satellite while the CE predicts a single peak. 
Fig. \ref{fig:dispersive_heat_map}(e) demonstrates this at $k=\pi/2$. 
There, the SC-CE predicts a quasiparticle and secondary peaks with positions, shapes and intensities in good agreement 
with exact results.
The CE, in contrast, simply predicts a broad Gaussian-like spectrum centered in an incorrect location.
At $k=\pi$ the SC-CE correctly predicts the very weak quasiparticle peak and the shape and intensity of the main satellite peak at $\omega = 2 \omega_0$. However, the spectrum is somewhat noisy and contains some small regions of negative spectral weight. 

Across the entire range of momenta, SC-CE correctly predicts the shapes and positions of the major peaks of the spectrum, and with the exception of $k=0$, is a vast improvement over the CE.
It appears that for more realistic dispersive models, the SC-CE is a very promising technique if one is accepting 
of physical defects, such as the prediction of small regions of negative spectral weight on the few percent level. It is possible these defects may be removed in higher-order versions of the theory, although the numerical solutions of the resulting equations may present a challenge. Future work investigating systems such as the electron gas should shed further light on the ability of the SC-CE to capture non-trivial physics in a wide range of distinct physical settings.

\section{Conclusion\label{sec:conclusion}}
In this paper we have presented a self-consistent cumulant expansion (SC-CE) for the description of the dynamics of a single electron coupled to a bath of phonons.  The models we have studied have direct relevance in polaron problems, and more broadly may be applied to purely electronic problems such as the interacting electron gas.  Furthermore, the work outlined here takes a step toward placing the cumulant approach to Green's functions on an equal footing with the standard Dyson-equation based formulation of field theory via the introduction of a hierarchy of self-consistent renormalized equations.  Via a link through the Dyson equation, we have connected the SC-CE at lowest order to a time-local self-consistent closure technique in the theory of classical turbulence. This connection gives insights into expectations for the successes and failures of the SC-CE.

We find that for the six-site Holstein model with a single phonon frequency at weak coupling, the SC-CE spectral function includes non-perturbative corrections beyond the second and fourth order bare cumulant expansions (CEs) and correctly describes the spectrum at all momenta while the CE is only accurate at $k=0$.
This is a consequence of the SC-CE capturing long time recurrences in the Green's function.
The SC-CE also introduces a small degree of non-physical negative spectral weight for some wave vectors.  More troubling, we find that the behavior of the SC-CE becomes unstable at long times.  It remains unclear the degree to which this reflects the need for better methods to solve the non-linear integro-differential equation that embodies the SC-CE, a failure of the SC-CE itself, or both.

We have studied the SC-CE for the Holstein model with a single phonon frequency and with dispersive phonons.  Increasing the system size of the non-dispersed Holstein model towards the thermodynamic limit partially suppresses the divergences prominent in the finite size lattice models. It is unclear if a truly infinite system would have divergences since the SC-CE equations becomes challenging to converge at later times. Using a broadening parameter larger than that employed in recent VD studies along with a long-time truncation of the SC-CE time evolution, comparisons may be made with the standard CE and exact numerical results.  We find that the SC-CE spectral function replicates the VD results well at all momenta and corrects many deficits of the CE.  Since natural internal damping of the decay of the Green's function occurs in more realistic models with a finite phonon dispersion, we turn to the dispersive Holstein model both as a means to assess the SC-CE in more realistic settings as well as to see if stability of the SC-CE is enhanced in such cases. Indeed, for the dispersive Holstein model, we find that no additional damping or truncation is needed, and the results greatly improve many of the features of the CE, albeit with the defect of a small degree of unphysical negative spectral weight.  

Given these findings, and the fact that the cost of the SC-CE is only $\approx O(N^2 (t_{\text{max}} / \Delta t)^2)$ where $\Delta t$ is the time step, we believe the SC-CE is promising and should be explored further.  An obvious next step would be the investigation of the homogeneous electron gas problem, where the screened Coulomb interaction would play the role of the damped phonon Green's function.  In addition, higher-order SC-CE schemes can be explored, although the hierarchy of self-consistent cumulant equations rapidly becomes extremely numerically involved.  These and other possibilities will be the subject of future investigations.

\section*{Acknowledgements}
The authors thank Prof. Janez Bon{\v c}a for providing the data from Refs. \citenum{Bonca2019} and \citenum{PhysRevB.103.054304}.
P.J.R. acknowledges support from the National Science Foundation Graduate Research Fellowship under Grant No. DGE-2036197.
I.S.D. acknowledges support from the United States Department of Energy through the Computational Sciences Graduate Fellowship (DOE CSGF) under Grant No. DE-FG02-97ER25308. 
This work used the Extreme Science and Engineering Discovery Environment (XSEDE), which is supported by National Science Foundation grant number ACI-1548562.\cite{xsede}
Specifically, it used the Bridges-2 system, which is supported by NSF award number ACI-1928147, at the Pittsburgh Supercomputing Center (PSC).
P.J.R. acknowledges the use of XSEDE resources under Grant No PHY-210108.
D.R.R. acknowledges support from NSF CHE-1954791.
\appendix

\section{Derivation of SC-CE\label{app:full_derivation_S}}
Closely following Dunn,\cite{dunn1975electron}
we start with the Green's function in the form Eq. \ref{eq:g_t_time_ordered_exp}, 
with the goal to evaluate the trace of the exponential to all orders.
Since there is a single electron on the lattice, we can 
easily determine all of the matrix elements by expanding the exponent as
\begin{equation}
\begin{aligned}
    &\mathcal{G}_k(t) = \mathcal{G}^{(0)}_k(t)  \bigg\langle \sum_{n=0}^{\infty}
    T\bigg[ \frac{(-i)^n}{n!}
    \sum_{\left\{q_j\right\}_{j=1}^n}
    \int_0^t d t_1 \dots \int_0^t d t_n  \\
    & \times g_{q_n, k} e^{-i t_n \epsilon_{k}} e^{i t_n \epsilon_{k + q_n}}A_{q_n}(t_{n})   \\
    & \times g_{q_{n-1}, k + q_n} e^{-i t_{n-1} \epsilon_{k + q_n}} e^{i t_{n-1} \epsilon_{k + q_n + q_{n-1}}}    A_{q_{n-1}}(t_{n-1}) \\
    & \dots \times g_{q_1, k + q_2 + \dots + q_n} e^{-i t_n \epsilon_{k + q_2 +  \dots + q_n}} e^{i t_n \epsilon_{k}}A_{q_1}(t_1) 
    \bigg]  \bigg\rangle.
\end{aligned}
\label{eq:product_standard_derivation}
\end{equation}
Eq. \ref{eq:product_standard_derivation} makes apparent that each application 
of $V$ has the effect of shifting all subsequent $k$-dependent terms by $q_i$. 
Here, $A_q(t) = b_q e^{-i \omega_q t} + b^\dagger_{-q} e^{-i \omega_q t}$, and $\sum_i q_i = 0$.
Note that we have kept notation general beyond the Holstein model such that the electron-phonon coupling constant depends on momentum.
This relationship can be represented with the differential operator 
$e^{q \cdot \frac{d}{d k}}$.  
We define a vertex operator $\Gamma_{qk}(t) = g_{qk} e^{-i t \epsilon_k} e^{q \cdot \frac{d}{dk}} e^{i t \epsilon_k}$
which allows us to rewrite the electron Green's function as 
an exponential
\begin{equation}
    \mathcal{G}_k(t) = \mathcal{G}_k^{(0)}(t) \left\langle T e^{-i \sum_q \int_0^t d\tau \Gamma_{qk}(\tau) A_q(\tau)}\right\rangle.
\end{equation}
The remaining trace is only over the phonons. 
Assuming that the phonons are purely harmonic, we can apply the standard
result
$
\langle e^{B} \rangle = e^{\frac{1}{2}\langle {B}^2 \rangle}
$, 
which is valid
for an operator $B$ which is a linear combination of $b$ and $b^{\dagger}$.
This expectation value is readily evaluated as $\langle T A_q(t)A_{q'}(t')\rangle = i D^{(0)}_q(t-t') \delta_{q,-q'}$.
This final trace further constrains the values that $q_i$ can take, so that only physical electron-phonon vertices are possible. 
This completes the first portion of the derivation, and we are left with an explicit form for $S$
\begin{equation}
\begin{aligned}
     S = 
     &-i\sum_{q}  
     \int_0^{t}d\tau \int_0^{\tau}d\tau'
     D_{q}^{(0)}(\tau-\tau') \Gamma_{-qk}(\tau) \Gamma_{qk}(\tau').
\end{aligned}
\end{equation}

To generate a self-consistent expression for $S$, we apply the Feynman operator ordering theorem.\cite{PhysRev.84.108}
The theorem states that given a functional of time dependent operators $F[\hat{A}(\tau),\hat{B}(\tau), \dots]$ where $\tau \in [0,t]$
and with a unitary operator of the form
$
    \hat{U}(\tau') = T\exp{\left(\int_0^{\tau'} d\tau \hat{P}(\tau) \right)}
$ then
\begin{equation}
\begin{aligned}
    &\hat{U}(t)F[\hat{A}(\tau),\hat{B}(\tau), \dots] = \\
    &T\left[ \exp{\left(\int_0^t d\tau\hat{P}(\tau)\right)}F[\hat{U}(\tau)\hat{A}(\tau)\hat{U}^{-1}(\tau), \dots] \right].
\end{aligned}
\end{equation}

Considering that we have expressed $\mathcal{G}_k(t)$ as a functional of differential operators, 
we choose $\hat{P} = -i\left(\epsilon_k + \frac{d}{d\tau}\phi_k(\tau) \right)$.
The scalar function $\phi_k(t)$ is only required to vanish at $t=0$
and the self-consistent approach presented in the main text corresponds to the 
choice $\phi_k(t) = -i C_k(t)$.

The transformed vertex functional can be simplified to
\begin{equation}
\begin{aligned}
     \bar{\Gamma}_{qk}(\tau) &\equiv \hat{U}(\tau) \Gamma_{qk}(\tau)\hat{U}^{-1}(\tau) 
     \\ &=  g_{qk} e^{- i \phi_k(\tau)}e^{q\cdot \frac{d}{d k}}e^{+i\phi_k(\tau)},   
\end{aligned}
\end{equation}
and the transformed $S$ is then given by
\begin{equation}
\begin{aligned}
    \bar{S} &= -i \int_0^t d\tau \left(\epsilon_k +  \frac{d}{d\tau}\phi_k(\tau)\right)
    \\
    &- i \sum_q \int_0^t d\tau \int_0^{\tau}d\tau' D^{(0)}_q(\tau-\tau')
    \bar{\Gamma}_{-qk}(\tau)\bar{\Gamma}_{qk}(\tau').
\end{aligned}
\end{equation}

Just as in the standard cumulant expansion, the cumulants of a given order can 
be computed by evaluating time-ordered powers of $S$
\begin{equation}
    C_k(t) = i\phi_k (t) + \sum_{n=1}^{\infty} \frac{1}{n!}T[\bar{S}^n]_c, 
\label{eq:cumulant_expansion_formal}
\end{equation}
where the notation $[...]_{c}$ denotes the cumulant of an operator, and $T[\bar{S}]_c = T[\bar{S}]$. Subsequent terms are given by
\begin{equation}
    T[\bar{S}^n]_c = T[\bar{S}^n] -\sum_{m=1}^{n-1}\frac{(n-1)!}{m!(n-m-1)!}T[\bar{S}^m]T[\bar{S}^{n-m}]_c.
    \label{eq:a8}
\end{equation}
Evaluating Eq. \ref{eq:cumulant_expansion_formal} with $\phi_k(t) = C_k(t)$ to leading order produces Eq. \ref{eq:sc_ce_cumulant}.

\section{SC-CE Recursive Expression
\label{app:sc-ce_recursive}}
Here we illustrate how the lowest-order SC-CE generates the exact second-order CE and approximations to all higher order CEs.
Considering only the lowest order in Eqs. \ref{eq:cumulant_expansion_formal} and \ref{eq:a8}, we can write
 Eq. \ref{eq:sc_action_gt} for the SC-CE as a self-consistent equation for the Green's function
\begin{equation}
\begin{aligned}
    \mathcal{G}_k(t) = \mathcal{G}_k^{(0)}(t)
    \exp\bigg(i & \sum_q \int_0^t d\sigma  \int_0^\sigma d\tau
    |g_{qk}|^2
    \\ \times &
    D^{(0)}_q(\sigma-\tau) 
    \frac{\mathcal{G}_{k-q}(\sigma) \mathcal{G}_{k}(\tau)}
{\mathcal{G}_{k-q}(\tau) \mathcal{G}_{k}(\sigma)}\bigg).
\end{aligned}
\end{equation}
In the weak coupling limit, we find an iterative solution starting with the bare Green's function.
The first iteration is straightforward and replicates exactly the second cumulant
\begin{align}
    \mathcal{G}^{(1)}_k(t) = \mathcal{G}^{(0)}_k(t)\exp(C_2(k,t)).
\end{align}
While the integrals for subsequent iterations may be analytically evaluated,
it is more instructive to consider the further iterations as functions of $C_2(k,t)$.

The second and all subsequent iterations now have exponential resummations
of the second-order cumulant itself. Explicitly,
\begin{equation}
\begin{aligned}
    \mathcal{G}^{(2)}_k(t) = & \mathcal{G}_k^{(0)}(t)
    \exp\bigg(-i  \sum_q \int_0^t d\sigma  \int_0^\sigma d\tau
    |g_{qk}|^2
    \\ \times &
    D^{(0)}_q(\sigma-\tau) 
    \mathcal{G}^{(0)}_{k-q}(\sigma - \tau) \mathcal{G}^{(0)}_{k}(\tau - \sigma) \\
\times & 
e^{C_2(k, \tau) + C_2(k-q, \sigma) - C_2(k, \sigma) - C_2(k-q, \tau)}
\bigg).
\end{aligned}
\end{equation}
The $n^{\text{th}}$ cumulant is defined as being proportional to the $n^{\text{th}}$ power of the coupling constant, so by expanding the exponent we find an approximation for all higher order cumulants
\begin{align}
    &\mathcal{G}^{(2)}_k(t) = \mathcal{G}_k^{(0)}(t)
    \exp\bigg(\sum_{n=0}^{\infty} \frac{-i}{n!}\sum_q \int_0^t d\sigma  \int_0^\sigma d\tau 
    \\ \nonumber \times &
    |g_{qk}|^2
    D^{(0)}_q(\sigma-\tau) 
    \mathcal{G}^{(0)}_{k-q}(\sigma - \tau) \mathcal{G}^{(0)}_{k}(\tau - \sigma) \\ \nonumber
\times & 
\left({C_2(k, \tau) + C_2(k-q, \sigma) - C_2(k, \sigma) - C_2(k-q, \tau)}\right)^n 
\bigg).
\end{align}
We can simplify this expression by introducing $C_{m}^{(n)}(k,t)$, which represents the approximation to the $n^{\text{th}}$ iteration of the $m^{\text{th}}$ cumulant.
This results in
\begin{align}
\mathcal{G}^{(2)}_k(t) = \mathcal{G}_k^{(0)}(t) \exp
\left(\sum_{n=1}^{\infty} C_{2n}^{(2)}(k, t) \right).
\end{align}
Importantly, because the second iteration produces cumulants of all orders, further iterations require expanding the exponent into moments.

After the second iteration, we can generalize to an arbitrary iteration:
\begin{align}
&\mathcal{G}^{(n+1)}_{\mathbf{k}}(t)  = \mathcal{G}_{\mathbf{k}}^{(0)}(t) 
\exp\Bigg( 
- i\sum_{\mathbf{q}}  
\int_0^t d\sigma
\int_0^\sigma d\tau
|g_{qk}|^2 \nonumber \\ 
& \times D^{0}_{\mathbf{q}}(\sigma - \tau) 
\mathcal{G}^{(0)}_{\mathbf{k}}(\tau-\sigma)
\mathcal{G}^{(0)}_{\mathbf{k}-\mathbf{q}}(\sigma-\tau)
e^{F^{(n)}(\mathbf{k},\mathbf{q}, \sigma, \tau)}
\Bigg).
\end{align}
Here, we introduce the notation:
\begin{align}
F^{(n)}(\mathbf{k},\mathbf{q},\sigma, \tau) &\equiv
F^{(n)}_2(\mathbf{k},\mathbf{q},\sigma, \tau) +F^{(n)}_4(\mathbf{k},\mathbf{q},\sigma, \tau) + \dots,   \\
F^{(n)}_i(\mathbf{k},\mathbf{q},\sigma, \tau) &\equiv C^{(n)}_i(\mathbf{k}-\mathbf{q},\sigma )-C^{(n)}_i(\mathbf{k}-\mathbf{q},\tau )
\nonumber \\ &
-C^{(n)}_i(\mathbf{k},\sigma )+C^{(n)}_i(\mathbf{k},\tau ).
\end{align}
The procedure for constructing the approximation to the $n^{\text{th}}$ iteration
of the $m^{\text{th}}$ cumulant is given by the moment expansion
\begin{align}
    \sum_{m=0}^{\infty} \lambda^{2m} W_{2m}^{(n)}(\mathbf{k}, \mathbf{q},\sigma,\tau) &  = 
    e^{\sum_{m=1}^{\infty} \lambda^{2m} F_{2m}^{(n)}(\mathbf{k},\mathbf{q},\sigma, \tau)},
\end{align}
where the first few moments are given by:
\begin{align}
    W^{(n)}_{0}(\mathbf{k}, \mathbf{q},\sigma,\tau) &= 1 \\
    W^{(n)}_{2}(\mathbf{k}, \mathbf{q},\sigma,\tau) &= F^{(n)}_2(\mathbf{k},\mathbf{q},\sigma, \tau)\\
    W^{(n)}_{4}(\mathbf{k}, \mathbf{q},\sigma,\tau) &= F^{(n)}_4(\mathbf{k},\mathbf{q},\sigma, \tau) \nonumber \\
    &+ \frac{1}{2}\left(F^{(n)}_2(\mathbf{k},\mathbf{q},\sigma, \tau)\right)^2.
\end{align}

Because each iteration introduces another power of $g^2$ to all of the approximated cumulants, and the $n^{\text{th}}$ order cumulant is defined as being proportional to $g^n$, the $n^{\text{th}}$ iteration cannot alter any of the approximated cumulants of order lower than $n$.
Because of this, and the fact that the approximate cumulant for a given order is an integral over lower order approximations, the iterative method is actually recursive.
Dropping the superscripts indicating iteration, the general form for the SC-CE approximation to the $n^{\text{th}}$ cumulant is 
\begin{align}
    C^{\text{SC-CE}}_{2n}&(\mathbf{k},t)  = 
    - i\sum_{\mathbf{q}}  
    \int_0^t d\sigma
    \int_0^\sigma d\tau
|g_{qk}|^2 D_{\mathbf{q}}(\sigma - \tau) \nonumber \\
\times & \mathcal{G}^{(0)}_{\mathbf{k}}(\tau-\sigma)
\mathcal{G}^{(0)}_{\mathbf{k}-\mathbf{q}}(\sigma-\tau)
W_{2n-2}(\mathbf{k}, \mathbf{q},\sigma,\tau).
\label{eq:sc-ce_recursive}
\end{align}

\section{Numerical VIDE Solution\label{app:numerics}}

Eq. \ref{eq:sc_action_gt} is not convenient for numerical calculations.
Instead, we define, $y_k(t) \equiv e^{C_k(t)}$ and solve directly for $y_k(t)$,
\begin{equation}
\begin{aligned}
    \frac{d y_k(t)}{d t} = &-i \sum_{q} \int_0^t d\tau |g_{qk}|^2 D^{(0)}_q(t-\tau) 
    \\
    & \times 
    e^{i(\epsilon_k - \epsilon_{k-q})(t-\tau)}
    \frac{y_k(\tau)y_{k-q}(t)}{y_{k-q}(\tau)}.
\label{eq:sc-ce-ykt}
\end{aligned}
\end{equation}

Numerical solutions to equation \ref{eq:sc-ce-ykt}
were carried out with a multi-step predictor corrector method
as derived by Linz.\cite{linz1969linear}
His method is general for first order
Volterra Integro-Differential Equations
(VIDEs) of the form
\begin{equation}
    y'(x) = F\left(x, y(x), \int_0^x dt K(x, t, y(t)) \right).
\end{equation}
The form of Eq. \ref{eq:sc-ce-ykt} implies that
$F$ can be written purely as a function of $\int_0^x dt K(x, t, y(t))$.
We used an Adams-Bashforth second-order predictor step and a 
third-order Adams-Moulton corrector step.
Following Linz, for the starting procedure of the Holstein model we applied the 
self-consistent Simpson's method iterated five times.\cite{linz1969linear}

\bibliography{references_II}

\end{document}